\documentclass[conference]{IEEEtran}
\IEEEoverridecommandlockouts

\usepackage{cite}
\usepackage{amsmath,amssymb,amsfonts}
\usepackage{algorithmic}
\usepackage{graphicx}
\usepackage{textcomp}
\usepackage{listings}
\usepackage[bookmarks=false]{hyperref}
\usepackage{makecell}
\usepackage{enumitem}
\usepackage[super]{nth}
\usepackage{tikz}
\usetikzlibrary{shapes,positioning,calc}

\def\BibTeX{{\rm B\kern-.05em{\sc i\kern-.025em b}\kern-.08em
    T\kern-.1667em\lower.7ex\hbox{E}\kern-.125emX}}
\begin{document}

\newlist{legal}{enumerate}{10}
\setlist[legal]{label*=\arabic*.}

\lstset{
    language=C++,
    breaklines = true,
    upquote = true,
    columns = flexible,
    basicstyle = \ttfamily,
    frame = single,
    keepspaces = true,
}

\title{Robust Resource Partitioning Approach for ARINC~653 RTOS
}

\author{\IEEEauthorblockN{Vitaly Cheptsov}
\IEEEauthorblockA{\textit{Ivannikov Institute for System Programming}\\\textit{of the Russian Academy of Sciences} \\
Moscow, Russia \\
cheptsov@ispras.ru}
\and
\IEEEauthorblockN{Alexey Khoroshilov}
\IEEEauthorblockA{\textit{Ivannikov Institute for System Programming}\\\textit{of the Russian Academy of Sciences} \\
Moscow, Russia \\
khoroshilov@ispras.ru}
}

\maketitle

%

\begin{abstract}
Modern airborne operating systems implement the concept of robust time and resource partitioning imposed by the standards for aerospace and airborne-embedded software systems, such as ARINC~653. While these standards do provide a considerable amount of design choices in regards to resource partitioning on the architectural and API levels, such as isolated memory spaces between the application partitions, predefined resource configuration, and unidirectional ports with limited queue and message sizes for inter-partition communication, they do not specify how an operating system should implement them in software. Furthermore, they often tend to set the minimal level of the required guarantees, for example, in terms of memory permissions, and disregard the hardware state of the art, which presently can provide considerably stronger guarantees at no extra cost. In the paper we present an architecture of robust resource partitioning for ARINC~653 real-time operating systems based on completely static MMU configuration. The architecture was implemented on different types of airborne hardware, including platforms with TLB-based and page~table-based MMU. Key benefits of the proposed approach include minimised run-time overhead and simpler verification of the memory subsystem.
\end{abstract}


\section{Introduction}

Multiprocess real-time operating systems are meant to provide different kinds of application software isolation, with time and resource partitioning being among of them. Time partitioning represents the way the operating system shares CPU time between the processes and resource partitioning defines how resource sharing is organised. System memory, consisting of RAM and device memory, is a primary asset which resource partitioning splits among the processes.


Traditionally real-time operating systems, such as FreeRTOS or RTEMS even with safety-critical profile~\cite{SafeRTOS178}, provided strong time partitioning guarantees but were oblivious to memory space partitioning. The primary cause for this separation is the inability of various microprocessors in the past or microcontrollers at present to provide this in hardware at reasonable performance. Nowadays there are many industries, where powerful microprocessors are used in combination with real-time operating systems, and airborne equipment is one of them.

The importance of memory space partitioning is vital for safety-critical systems as it is a requirement for a possibility to transition from ``multiple applications on multiple hardware'' approach to ``multiple applications on single hardware''. This transition happened in avionics as part of the migration to IMA~\cite{IMA} architecture and resulted hardware costs and weight reduction, as well as the ability to build more complex software with better flexibility.

An operating system with explicit memory space partitioning guarantees that software errors like memory corruptions in one process cannot negatively affect other processes. This can be achieved by the use of virtual memory spaces, a feature present in most microprocessors of the past 20~years. For general purpose operating systems virtual memory use is an established practice, but when it comes to robust resource partitioning, an operating system is meant to provide even stronger guarantees. For example, memory allocations in one process shall not lead to memory allocation failures in another. This quality is much less common and is generally found only in specialised environments, such as safety-critical hard real-time systems.

In this paper we cover the architecture and design of robust memory space partitioning for an ARINC~653~\cite{Arinc653P1} operating system (sections~\ref{subsec:arch} and ~\ref{subsec:impl}), which not only qualifies to execute all memory partitioning requirements imposed by the ARINC~653 standard, but also simplifies verification (section~\ref{subsec:verif}), provides high performance for the memory subsystem, and reduces the complexity of worst-case execution time calculation in terms of memory usage (section~\ref{subsec:perf}). The key design decisions in the paper were evaluated in the existing operating system framework CLOS, developed at ISP\,RAS, and confirmed working on existing airborne equipment.

\section{State of the Art} \label{subsec:stateofart}

\emph{Robust partitioning} concept is in the core of ARINC~653 specification~\cite{Arinc653P0}. Originally it appeared in DO-248~\cite{Do248} standard and was later extended for multicore support in CAST-32A~\cite{Cast32A} position paper. ARINC~653 implements this concept as a part of \emph{Robust resource partitioning} and \emph{robust time partitioning}. Taking these documents into consideration altogether, the system with robust partitioning can be described as the one implementing the following properties:

\begin{itemize}
\item Software partitions cannot contaminate the storage areas for the code, I/O, or data of other partitions.
\item Software partitions cannot consume more than their allocations of shared resources.
\item Failures of hardware unique to a software partition cannot cause adverse effects on other software partitions.
\item No software partition is permitted to consume more than its allocation of execution time on the CPU cores on which it executes, irrespective of other partition activity or inactivity on any CPU cores.
\end{itemize}

While good to have, the requirements presented only define the constraints the implementation should operate in and do not define the operating system architecture to be used to achieve them. For example, they do not provide the answers to many design questions, e.g.:

\begin{itemize}
\item How kernel-backed objects like inter-partition communication services such as queuing or sampling ports shall be allocated?
\item How shall memory subsystem guarantee that software partitions can coexist in the specified configuration?
\item How one should verify that software partitions are actually isolated memory-wise from each other?
\item How long will memory access take within the application in regards to memory subsystem functioning on the particular hardware if virtual memory is involved?
\end{itemize}

Software using ARINC~653 APIs is named APEX software and objects provided by ARINC~653 APIs are named APEX objects. APEX software is normally delivered as part of integration projects --- a preconfigured set of an operating system kernel, system libraries, and application software partitions for particular purpose and equipment. This approach enables an operating system developer to take many decisions at early steps of the integration process. Therefore, the generally accepted solution to enable robust resource partitioning found in multiple commercial ARINC~653 operating systems, like VxWorks~653 or MACS2, is to use unique address spaces created with the help of virtual memory per each software partition and perform shared resource allocation statically at project build time. However, many ARINC~653 operating systems still perform dynamic resource allocation, including address space build-up, which means the answers to the questions stated above are given late in runtime. Some systems may even rebuild address spaces during application execution, which can negatively affect memory access performance and makes address space management subsystem verification a rather convoluted task.

\section{Robust Resource Partitioning Architecture} \label{subsec:arch}

Let's define robust partitioning requirements in a stricter manner than ARINC~653 to design the robust resource partitioning in a RTOS:

\begin{enumerate}
\item All kernel and software partition resources, including memory regions and CPU time, shall be statically defined by the configuration set at project build time.
\item All memory regions should have minimum access permissions required for normal kernel and software partition functioning as permitted by the underlying hardware.
\item Software partition execution time shall be limited by the configuration and shall not be exceeded.
\item Software partition execution shall not be interrupted by any means but to perform partition switching or executing actions requested by the partition itself.
\end{enumerate}

Items 1 and 2 shall be provided by robust resource partitioning and require a way to model integration project memory. Let's design a robust resource partitioning architecture that will execute these requirements based on the following principles:

\begin{enumerate}
\item Static resource management.
\item Static memory layout.
\item Static MMU configuration.
\end{enumerate}

\subsection{Static Resource Management}

To achieve static resource management in an ARINC~653 based operating system one needs to ensure that all the APEX objects are reserved at build time. These objects are split into two categories:

\begin{itemize}
\item Shared objects --- queuing or sampling ports, service access points, logbooks, etc.
\item Private objects --- processes, buffers, blackboards, mutexes, etc.
\end{itemize}

For shared objects the ARINC~653 standard requires the developer to specify the amount and parameters of each shared object in a declarative format, generally the XML configuration. To achieve static resource management for shared objects it is enough to translate these properties into build-time requirements.

For private objects the ARINC~653 standard does not always provide an established way to specify their properties, but as a matter of fact all ARINC~653 based operating systems do have this level of configuration in an implementation specific format. For example, some operating systems may explicitly provide a way to set the buffer count and some may only give an ability to set the overall heap size from which these objects can be allocated.

In the rest of this paper we take static resource management presence for granted in an APEX-based operating system.

\subsection{Static Memory Layout}

To achieve static memory layout let's define a \emph{memory block} as a means to describe the integration project requirement in system memory with the following attributes:

\begin{itemize}
\item Owner --- entity setting the memory block requirement within the integration project, can be a software partition or an OS kernel itself.
\item Logical name --- string used by an entity to gain access to the memory region described by the memory block.
\item Address --- address (generally virtual) used by an entity to perform memory access.
\item Physical address --- memory region address in system memory. If address does not match physical address (is virtual), address translation is performed by the hardware.
\item Size --- memory region size.
\item Access permissions --- memory access permissions limiting access operations to the memory region, can be read, write, or execute.
\item Cache policy --- memory access discipline hiding architecture-dependent way to configure memory region behaviour, can be ``normal'', ``IO'', ``normal+coherent'' and similar, which can vary based on memory type and the need of memory sharing.
\item Alignment --- minimal address alignment for the memory block.
\item Physical contiguousness --- whether memory region should be contiguous in system memory (for DMA, IO).
\end{itemize}

Following this model we can describe any integration project memory requirements as a combination of \emph{memory blocks} with partially defined attributes. The attributes can be partially defined when specific values are not relevant for project functioning and can be chosen automatically based on hardware and software capabilities. For example, a pool-based heap in a software partition should have predefined size and access permissions but can have arbitrary virtual and physical addresses. Similarly software partition code segment may have an arbitrary physical address, but may need fixed virtual address.

If system memory is split into memory blocks at build time and is not otherwise dynamically distributed, what remains to be done for robust partitioning in an ARINC~653-based system is as follows:

\begin{enumerate}
\item Assign all the missing attributes for the memory blocks.
\item Generate a compatible hardware configuration.
\item Perform verification of the generated data.
\end{enumerate}

\begin{figure}[h]
	\centerline{\includegraphics[scale=0.16]{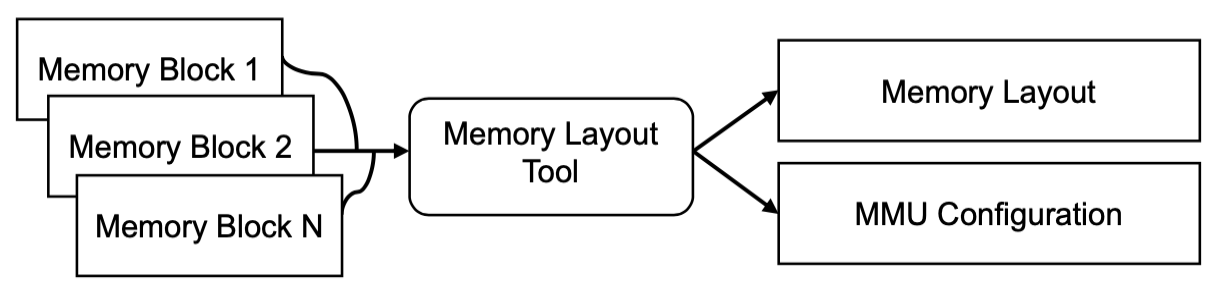}}
	\caption{Memory layout process}
	\label{fig:MemLayout}
\end{figure}

Let's call the process of assigning these attributes the \emph{memory layout process} and name a prefilled set with all the attributes assigned in correspondence to each other and platform capabilities the \emph{memory layout}. While the memory layout process can be performed by hand, a dedicated \emph{memory layout tool} can be developed to generate the memory layout automatically as shown on fig.~\ref{fig:MemLayout}.

\subsection{Static MMU Configuration}

After the memory layout is created for an integration project, what remains to be done is to adapt this memory layout for the particular hardware setup and to confirm whether it is possible or not to run this project on this hardware. While usually MMU configuration is computed within RTOS on the target hardware, our approach is to precompute it by memory layout tool. Figure \ref{fig:MemLayout} names this configuration \emph{MMU configuration} by the name of the memory management unit device responsible for virtual memory mapping.

While each platform has its own way to configure virtual memory, there are two major groups MMU devices can be split into:


\begin{itemize}
\item \emph{Manually configurable TLB MMUs}. For them virtual memory configuration is defined by the TLB unit (Translation Lookaside Buffer). Each TLB unit can be directly configured through the driver by writing to a fixed set of TLB entries (sometimes named segments). Each TLB entry can map a variably sized memory region, usually power-of-two, with unique memory permissions. TLB entries are stored in registers, and their number varies from tens to hundreds in modern CPUs. The TLB entry choice algorithm depends on the target architecture and can be fully associative, partially associative, or even have a special formula. The concept is typical for ARMv7-M/R, Armv8-M/R AArch64, PowerPC, and MIPS CPUs~\cite{E500MCRM}.
\item \emph{Page table MMUs}. For them virtual memory configuration is defined by trees (nested tables) in system memory indexed by bitfields of the virtual address. Memory mapping happens in fixed size areas named pages. Page table tree depth generally does not exceed 5 and can sometimes be shortened by consolidating lowest levels to map areas in larger pages. TLB units are also present in these MMUs, but they have no direct control and are used to optimise memory translation performance by caching recent address translations. The concept is typical for ARMv7-A, AArch64, RISC-V, and x86 CPUs~\cite{IntelSDM}.
\end{itemize}

Precomputing the MMU configuration before operating system launch gives an opportunity to take longer time on choosing optimal configuration. This benefit is unavailable with the ``regular'' approach as calculating the configuration on target hardware at boot time is constrained by reboot timing requirements and performing corrections during application partition execution may strongly increase memory latency. If done right, it may not only improve operating system performance, but also help to avoid unnecessary software interruptions as per item 4 in the robust partitioning requirements and gradually improve worst-case execution time (WCET).

\section{Static MMU Configuration Implementation} \label{subsec:impl}

Let's design the implementation separately for different types of MMUs.

\subsection{Fixed TLB entry approach} \label{subsec:tlbent}

Among the common pieces of hardware found in aerospace embedded solutions are the CPUs built on top of the PowerPC architecture, namely the QorIQ family, e.g. e500-series CPU cores from NXP. The e500mc memory management unit represents a pair of manually configurable TLBs~\cite{E500MCRM}:

\begin{itemize}
\item TLB0 has a 4-way associative structure, allowing to map 4 KB regions based on the virtual address tag. The typical amount of TLB0 entries is 4x128.
\item TLB1 has a fully associative structure, allowing to map a power-of-two memory region   with its virtual and physical addresses aligned by its size. The typical amount of TLB1 entries is 64.
\end{itemize}

There are several more CPU cores with similar MMU architecture, with some of them suitable for aerospace industry needs:


\begin{itemize}
\item MIPS CPUs starting with R3000 series and various clones from different vendors. These CPUs are special by their lack of N-way associative mode, which is compensated by splitting TLB entries into halves and providing fixed memory mapping for kernelspace. The typical amount of doubled TLB entries is 64.
\item PowerPC 470S CPUs, such as NTC Module 1888TKh018. These CPUs are special by their 4-way associative structure with a hash function used to select TLB entries. Their TLB entries similarly provide an ability to map power-of-two memory regions, but besides that support cacheability via ITLB and DTLB units in LRU mode. The typical amount of main TLB entries is 4x256~\cite{PPC476FP}.
\item While ARMv7 and ARMv8 CPUs with M and R profiles, such as Cortex-M4, do not support virtual memory, they have a memory permission unit (MPU) device built on similar principles. Their MPUs allow setting permissions in power-of-two memory regions with these regions being possibly split into several fixed length subregions. Access to each subregion can be programmatically disabled. The typical amount of regions is 24~\cite{CORTEXM4}.
\end{itemize}

General purpose operating systems often have high memory consumption and sometimes even involve overcommit memory management policies, and as a result they can only utilitise these TLB MMUs as a cache for a programmatically controlled page table. When application software performs memory access currently not present in the TLB an interrupt happens, and an algorithm called TLB refill swaps the least recently used TLB entry with a new one from the page table.

Let's call this approach a naïve approach to perform address space management with TLB MMUs, and understand why it does not work well for safety-critical hard real-time operating systems, such as ARINC~653 systems:


\begin{itemize}
\item WCET for address translation is either assumed to be equal to the WCET of the TLB-refill interrupt handling or needs to be calculated in a non-trivial way, still costing microseconds or more in favoured locations.
\item TLB refill code is often reasonably complicated and introduces additional risks into address space management at RTOS operation. Due to this code presence both static and dynamic verification of the address space management subsystem get more challenging.
\end{itemize}

To circumvent these downsides we at least need to avoid TLB refill interrupts at application execution time. For that purpose we propose \emph{Fixed TLB entry address space management} method. In this method each partition is statically assigned a list of fixed TLB entries, which are written at once at address space switching time. Since address space switching in ARINC~653 operating systems generally happens at the end of the schedule window, the suggested alternative successfully lacks the downsides of the naïve approach:

\begin{itemize}
\item WCET for address translation now only depends on the TLB implementation in hardware, so the value will be both low and well-known.
\item Moving TLB management to specific places and pre-calculating the written TLB entry sequences reduces the total amount of states address space management subsystem can be in, easing verification and simplifying driver internals.
\end{itemize}

Despite method simplicity, it does not have significant downsides for the segment ARINC~653 RTOS operate in. To prove this, let's consider three main areas for potential issues: memory usage, address space switching time, and most importantly memory layout existence.

TLB entry sequences used to perform software partition address space switching need to be stored in memory. While it may seem that the overall system memory usage may increase, for all the three, PowerPC e500-series, PowerPC 470s, and MIPS R4000, just five 32-bit values are enough to natively encode one TLB entry. As switching to this method also avoids the need in software page tables, we discovered no memory consumption increase. Furthermore, the benefit of having these values pre-calculated is the ability to store them in read-only memory for extra protection from memory corruptions.

Address space switching time may increase in absolute values, but that benefit of the naïve method exists at the cost of WCET increase for all memory accesses. Formally speaking, updating TLB configuration at once in a loop over the TLB sequence at address space switching cannot take more time than handling all TLB refill interrupts, which may happen in the worst scenario. This is because the amount of written TLB entries will be about the same, if not less in the suggested method, but the naïve method will also have to perform the interrupt handling. Taking this into account one will also have to increase  the window lengths in the schedule in order to provide enough spare time for TLB refill interrupt handling.

On the other side, most TLB memory management units support process context identifiers, enabling them to have TLB entries written in the registers but not being active till the particular task is switched to. If TLB size is large enough to account for more TLB entries, the amount of overwritten TLB entries at each address space switching routine may be considerably reduced.

As for memory layout existence, the suggested method obviously makes certain memory layouts impossible to configure on hardware. As a result many vendors consider it the design risky due to the chance of such layouts to be needed by the integration projects. This risk essentially comes down to whether the amount of available TLB entries in hardware is enough to map the \emph{incompatible} memory blocks, i.e. blocks with unique permissions and cache policy or disjoint addresses. The important compromise with this method is to map only one software partition and the operating system kernel within a single sequence of TLB entries at a time (per address space). Kernelspace and userspace code and data segments will give 4 used TLB entries as a minumum. Even the absolutely low end NXP QorIQ P1010 with PowerPC~e500v2 core and just 16 TLB entries leaves 12 remaining entries for stacks, object pools and IO. Our experience shows that in real world examples it is even possible to dedicate a separate TLB entry per process stack on this low end hardware, let alone more current hardware, which has several times more TLB entries.

To sum it up, the suggested method preserves the robust resource partitioning properties defined in section~\ref{subsec:arch}, and enables TLB-based CPUs to perform efficient memory address translation at average and in the worst case scenario. Fixed TLB entry sequences reduce the amount of states address space management subsystem can be in, and simplify its verification.

\subsection{Page tables with deterministic timing} \label{subsec:pt}

Common CPUs found in industrial grade SBCs are often built on top of more modern architectures, such as AArch64 and RISC-V. These architectures have their MMUs designed based on the page table model. Although historically ARM-based architectures have long had R-profile for hard real-time and safety critical systems, the lack of virtual memory support in R-profile CPUs makes them impractical for many segments of IMA equipment. As a result today it is a matter of fact to have CPUs with page table MMU architecture in safety-critical airborne equipment.

To translate a virtual address into a physical one in the page table model, the processor makes up to N sequential system memory accesses corresponding to the elements of the page table tree for a given virtual address. N is defined by hardware and is the maximum number of nesting levels in the page table tree. For fast memory, for example, LPDDR4, used in systems with a non-aggressive environment, such accesses will take tens of nanoseconds, which is reasonably acceptable, but for slow radiation-resistant RAM such delays can be hundreds of nanoseconds and more.

In order to optimise the translation process page table based MMUs are equipped with a TLB unit. This device caches translated addresses in much faster memory and is called first before performing the usual translation process involving system memory accesses. The unfortunate downside of this design is that it is not possible to directly manipulate the contents of the TLB unit, which are instead automatically filled based on the address usage frequency with an LRU algorithm. Being able to only directly erase TLB contents makes predicting the address translation time a non-trivial task, which needs to be solved for safety-critical systems.

If we think of this task from the point of a naïve model of address space management for TLB-based MMUs described in section \ref{subsec:tlbent}, we will find many similarities. Both models have address space configuration stored in system memory, both models utilise TLB units as a cache, and both models involve a TLB refill process. The only notable difference between the two is that for page table MMUs the TLB refill process is performed by the hardware and is fully automatic. Still, while hardware TLB refill makes the first address translation happen much faster, assuming its time to be equal to N system memory accesses is unacceptable for most WCET calculation scenarios.

While it is definitely not possible to redesign the TLB unit logic in these CPUs, we can still significantly increase the level of determinism during address translation. To do this we present an approach to use \emph{page tables with deterministic timing} built on the following principles:

\begin{enumerate}
\item Using global address space identifier for all kernel pages.
\item Performing TLB cleanup for non-global address space identifiers at address space switch time.
\item Using faster system memory (e.g. SoC internal RAM) to store page tables when available.
\item Minimising page count based on the particular TLB spec, i.e. the amount of TLB slots for each page level.
\item Pre-filling (warming up) TLB at address space switching time by generating memory accesses.
\end{enumerate}

Let's have a closer look on each principle.


Most MMUs have a concept of local address space identifiers (PID, PCID, etc.)~\cite{IntelSDM}, which can be transparently used along with the virtual address tag at TLB lookup times. This design decision allows to store virtual address from different address spaces without conflicts and to perform partial TLB cleanup, removing entries for a particular address space upon request. If a local identifier corresponds to an address space of a particular software partition, a global identifier is a special identifier that is common across all address spaces. By using local and global identifiers as per items 1 and 2 one can improve address space management determinism as follows:

\begin{itemize}
\item Using global identifier for all kernel pages allows to ensure that kernel TLB entries are not evicted at address space switching and therefore increases the overall kernel performance.
\item Using unique local identifiers for software partitions allows to independently store software partition TLB entries. Since ARINC~653 limits the total amount of supported partitions by a fixed value, it is often possible to have one-to-one matching between software partitions and local identifiers.
\item Performing TLB entry cleanup for a particular local identifier at address space switching is reasonable when the total amount of available TLB slots is less than the potential amount of TLB entries for all partitions in an integration project. Doing this lets the partition behave in a similar way regardless of the window in the schedule.
\end{itemize}

Most SoCs offer a small amount of fast internal RAM, generally used for platform initialisation by the bootloader. This memory commonly has much lower delays than external RAM and may be unused by software after the chip is initialised. Embedded operating systems often have a close and flexible connection with the components executed early in the boot chain and under certain conditions the internal memory in whole or in part can be left at RTOS kernel disposal. Given this is the case, such memory areas can be used to store critical data with page tables being one of them. For certain CPUs the benefits of using the internal RAM are not just the performance gains but also better reliability, as external memory chips and the interconnect can be exposed to an aggressive environment.

Last but not least comes the TLB entry locality. This task is uneasy to start with, because some vendors do not provide much information about the TLB structure, but given the necessary documentation from the chip maker or taking time to perform hardware tests, all the necessary knowledge can be extracted. Assuming we know how many entries can a TLB unit hold, there opens a possibility to never evict a TLB entry during software partition execution window once it was written. For this we should limit the total amount of pages in each software partition and the kernel to be less or equal to the amount of TLB entries available in the TLB unit at memory layout stage, and then simply let it execute. Given hundreds of TLB entries in modern CPUs (e.g. 512 or 1024 in common AArch64 cores like Cortex-A53 or Cortex-A55), this task is often perfectly doable for ARINC~653 integration projects.

Moving with this task further, as per item 5 we would like to move the first address translations requiring system memory access to the address space switching point. Since most second level TLBs have a unified architecture, i.e. they do not separate TLB entries for reads, writes, or code execution, any memory access will cache the virtual address in the TLB unit. With this in mind it is enough to perform a sequence of reads one per page page to preload TLB contents, essentially warming it up and avoiding extra delays during software partition execution.

To sum it up, the suggested method preserves the robust resource partitioning properties defined in section~\ref{subsec:arch}, and makes address space subsystem based on page table MMUs more deterministic and performant. Under certain conditions it may be possible to entirely avoid TLB refill and thus reduce memory translation latency during software partition execution to the minimum provide by TLB hardware.

\section{Verification} \label{subsec:verif}

To ensure the correct functioning of the address space management subsystem in an integration project one must ensure that the following properties are true:

\begin{enumerate}
\item System memory map corresponds to the actual hardware and platform initialisation software used.
\item Requested memory requirements correspond to actual functional requirements of the integration project.
\item Provided address spaces correspond to memory requirements and are isolated from each other.
\item Address spaces are invariant to memory requirements throughout the entire integration project execution time.
\end{enumerate}

Verification of items 1 and 2 can be performed manually. For practically any ARINC~653 RTOS and integration project this involves the DO-178C~\cite{Do178} process and is out of the scope of this paper. An architectural decision described in section~\ref{subsec:arch} and implemented in sections~\ref{subsec:tlbent}~and~\ref{subsec:pt} allows us not to consider any specific verification procedures for item 4, because address space layouts are static and thus invariant. As for item 3, with the help of automated tooling we can not only check the configuration properties, but also relax the certification requirements for the memory layout tool as it becomes far less critical.

There are two major types of errors that can exist in the address space management subsystem:

\begin{itemize}
\item Configuration errors --- errors caused by an incorrect memory layout, which for some reason does not correspond to the requirements.
\item Implementation errors --- errors caused by software malfunctioning (e.g. memory management unit driver bugs) or hardware errata.
\end{itemize}

Let's consider a combined approach of static and dynamic verification methods to detect these errors.

\subsection{Static Memory Layout Verification}

Let's consider a schema described on fig.~\ref{fig:MemLayout} from section~\ref{subsec:arch}. A set of memory blocks along with the system memory map data is passed to the memory layout tool in order to obtain a memory layout and an MMU configuration. In this sense the memory layout tool can be viewed as a translator from memory blocks to memory layout. In order to verify the resulting memory layout correctness we can then prove that the input data is correct and the output data corresponds to the input data.

\begin{figure}[h]
	\centerline{\includegraphics[scale=0.12]{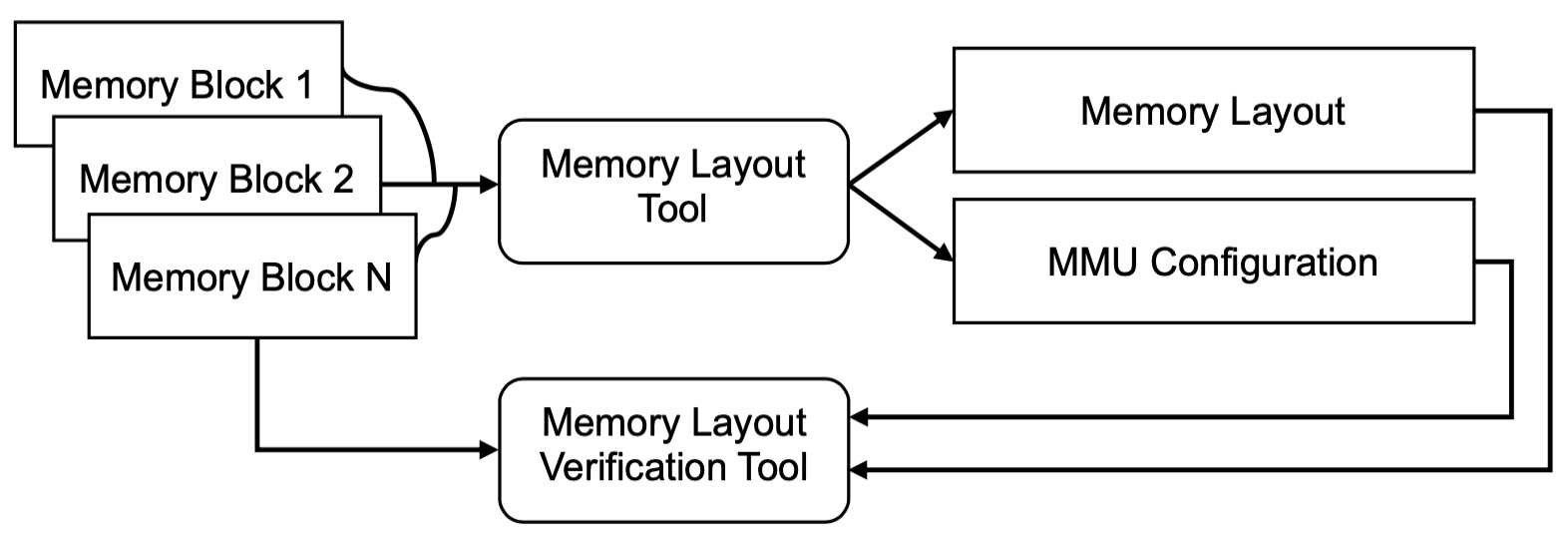}}
	\caption{Static memory layout verification}
	\label{fig:StaticVerif}
\end{figure}

Let's add a \emph{memory layout verification tool} to the process as shown on fig.~\ref{fig:StaticVerif}. With this change instead of a single step memory layout process we get a three-step solution:

\begin{enumerate}
\item Memory layout verification tool checks that memory block requirements are not malformed.
\item Memory layout tool performs the memory layout process.
\item Memory layout verification tool checks that the memory layout corresponds to the memory block requirements in both the architecture-independent part (the memory layout itself) and the architecture-dependent part (the MMU configuration).
\end{enumerate}

\subsection{Dynamic Memory Layout Verification}

Verifying memory layout at runtime allows to detect various implementation errors such as 
incorrect memory layout interpretation by the MMU driver or logic errors in the driver itself. The room for these errors can be significantly reduced by choosing the MMU configuration data format as close to the target hardware as possible, for example, the TLB entries can be stored as precomputed register values and page tables can be pre-built. However, there always is a possibility of an error in the static verification tool or even at an earlier step, when the properties of the target system are analysed. For example, the MMU characteristics on actual hardware may slightly differ from the ones written in the specification provided to the memory layout tool, and thus the error will remain undetected during static verification.

\begin{figure}[h]
	\centerline{\includegraphics[scale=0.12]{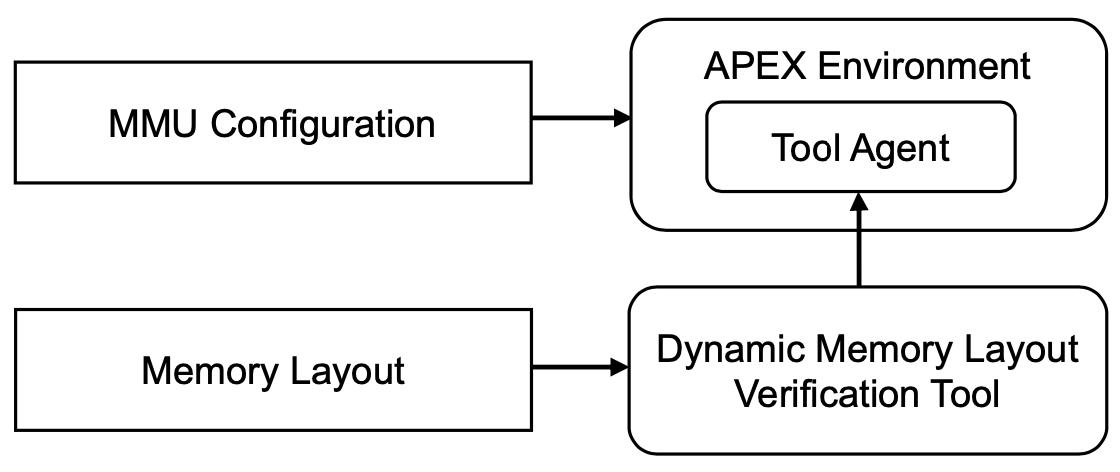}}
	\caption{Dynamic memory layout verification}
	\label{fig:DynamicVerif}
\end{figure}

While it may seem straightforward to perform this kind of verification by running the integration project on the target device, many errors cannot be detected this way. For example, if the software partition does not try to access kernel memory we may not know about the privilege violation issue till this partition is compromised and an attack happens. To circumvent this limitation let's consider a dedicated structure with an agent inside the operating system from fig.~\ref{fig:DynamicVerif}.

The \emph{tool agent} implements the following functionality:

\begin{itemize}
\item Establishing a communication channel with the \emph{dynamic memory layout verification tool} (runs externally).
\item Performing memory access at specified address and size with selected software partition or kernel permissions.
\item Reporting memory access violation exceptions over the communication channel including the information about memory violation type, address, etc.
\item Recovering from the memory access violation exceptions.
\end{itemize}

The \emph{dynamic memory layout verification tool} then implements the following checks:

\begin{enumerate}
\item ``Positive'' memory accesses to memory blocks according to the memory layout, i.e. try to read readable memory, write writable memory (and restore afterwards), execute executable memory.
\item ``Negative'' memory accesses to memory blocks and outside areas.
\end{enumerate}

If a ``positive'' memory access fails or a ``negative'' memory access succeeds then a potential bug is found.


\section{Performance evaluation} \label{subsec:perf}

We successfully evaluated the suggested approach, including the verification toolset, in an ARINC~653 based RTOS developed at ISP~RAS~\cite{jetos} on multiple SBCs on ARMv7-A, AArch64 (ARMv8-A), MIPS32, and PowerPC architectures.

In order to measure performance we took an avionics computer with 1.2 GHz NXP P3041 QorIQ CPU on PowerPC architecture and 1200 MT/s DDR3 memory and performed a number of measurements of TLB operations and interrupt handling in single core mode with caching enabled:

\begin{itemize}
\item 22 TLB ops --- up to 1280~ns, 770~ns average.
\item 45 TLB ops --- up to 2454~ns, 1542~ns average.
\item 64 TLB ops --- up to 3520~ns, 2133~ns average.
\item Synchronous interrupts --- 618~ns at lowest.
\end{itemize}

Following the numbers, even the pessimistic estimate shows that implementing the suggested approach in the operating system saves over 650~ns WCET per memory access, which already is an order of magnitude greater than already translated memory access time.

A potential increase in address space switching time does not seem to be significant, and at very least a magnitude lower than a sum of lazy TLB refill operations at runtime, which can now be deduced from the partition window duration.


\section{Conclusion}

In this paper we proposed an approach to implement robust resource partitioning in an ARINC~653 RTOS. The key principles of the method are static resource management, static memory layout, and static MMU configuration. The last two principles are not known to be used in any RTOS with time and space isolation, be that an ARINC~653 RTOS or a RTOS with comparable API complexity.

Chosen principles simplify the verification process of the RTOS address space management subsystem. In the paper we described two distinct approaches to perform address space management subsystem built on these principles.

The method and the verification toolset were successfully implemented in an ARINC~653 based RTOS developed at ISP~RAS for multiple target architectures. The performance analysis showed a significant reduction of WCET for memory accesses, generally allowing to reduce partition window durations in the schedules.


\section*{Acknowledgements}

ISP~RAS and ISP~RAS~Open committee for review and comment. All the members of the hard real-time operating systems development and verification group at ISP~RAS for their invaluable assistance. Andrey Tsyvarev, Maxim Doledenok, and Sophia Zelenova in particular.


\end{document}